%% file: kdd.tex
\documentclass[sigconf]{acmart}

\setcopyright{none}
\renewcommand\footnotetextcopyrightpermission[1]{}
\usepackage{balance}
\usepackage{booktabs} 
\usepackage{listings}
\usepackage{color}
\usepackage[skip=2pt]{caption}
\usepackage{subcaption}
\usepackage{cleveref}
\usepackage{textcomp}

\def\BibTeX{{\rm B\kern-.05em{\sc i\kern-.025em b}\kern-.08emT\kern-.1667em\lower.7ex\hbox{E}\kern-.125emX}}

\DeclareCaptionFormat{captionformat}{\fontsize{7}{8}\selectfont#1#2#3}
\definecolor{dkgreen}{rgb}{0,0.6,0}
\definecolor{gray}{rgb}{0.5,0.5,0.5}
\definecolor{mauve}{rgb}{0.58,0,0.82}

\mathchardef\mhyphen="2D

\begin{document}
\title{Managing Diversity in Airbnb Search}
\author{Mustafa Abdool, Malay Haldar,  Prashant Ramanathan, Tyler Sax, Lanbo Zhang}
\author{Aamir Mansawala, Shulin Yang, Thomas Legrand}
\affiliation{%
  \institution{Airbnb Inc.}
}
\email{moose.abdool@airbnb.com}

\renewcommand{\shortauthors}{Mustafa Abdool et al.}

\begin{abstract}

One of the long-standing questions in search systems is the role of diversity in results. From a product perspective, showing diverse results provides the user with more choice and should lead to an improved experience. However, this intuition is at odds with common machine learning approaches to ranking which directly optimize the relevance of each individual item without a holistic view of the result set. In this paper, we describe our journey in tackling the problem of diversity for Airbnb search, starting from heuristic based approaches and concluding with a novel deep learning solution that produces an embedding of the entire query context by leveraging Recurrent Neural Networks (RNNs). We hope our lessons learned will prove useful to others and motivate further research in this area.
\end{abstract}

%
%



\maketitle

\input{kddbody-conf}

\bibliographystyle{ACM-Reference-Format}

\balance 
\bibliography{kdd-bibliography}

\end{document}

%% file: kddbody-conf.tex
\section{Introduction}
Airbnb is a two sided marketplace with the ultimate goal of connecting hosts, who provide places to stay, with prospective guests from around the globe. The search ranking problem at Airbnb is to rank the hosts' properties, referred to as \textit{listings}, in response to a query which typically consists of a location, number of guests and checkin/checkout dates. Listings themselves possess a rich taxonomy ranging from entire homes to boutique hotels and an ideal ranking model would be able to determine exactly which listings are most relevant by taking into account both guest and host preferences.
    
Airbnb's search ranking algorithm has undergone major changes since its inception --- the most recent being the transition to deep learning detailed in ~\cite{kdd19}. However, the question of listing diversity in search results had existed long before the advent of deep learning at Airbnb. Since the earliest days of the ranking team, we had theorized that managing diversity could lead to a better search experience by providing more choices to users as demonstrated in other e-commerce applications such as Amazon's \textit{Interesting Finds} product ~\cite{teo2016adaptive}.

Over the years, we have explored various techniques to manage diversity, most of which evolved in parallel to the main ranking model. These solutions range from heuristic techniques inspired by the classic Information Retrieval (IR) literature to complete end-to-end deep learning solutions that leverage Recurrent Neural Networks (RNNs). In this paper, we give an account of both the successes and failures on our journey to understand the role of diversity in search along with important lessons learned along the way.

\subsection{Ranking Problem Formulation}

The search ranking problem is modelled as one of \textit{pairwise preference} which is a common approach found in the Learning To Rank literature ~\cite{cao2007learning}. Each training instance is a pair of a booked listing and a non-booked listing for a given query and user. The ranking model itself is a Deep Neural Network (DNN) and produces a score given some listing, query and user features:

\begin{equation}
\begin{aligned}
S_{booked} &= F_{\theta}(L_{booked}, Q, U) \\
S_{unbooked} &= F_{\theta}(L_{unbooked}, Q, U) \\
PairwiseLoss &= CE(S_{booked} - S_{unbooked}, 1)
\end{aligned}
\label{pairwise_eqn}
\end{equation}

Where:

\begin{itemize}
    \item $F$ represents the function computed by the DNN (with parameters denoted by $\theta$)
    \item $L_{booked}$ and $L_{unbooked}$ denote the listing feature vectors for the booked and non-booked listings respectively
    \item $Q$ and $U$ denote query and user features respectively
    \item $S_{booked}$ and $S_{unbooked}$ denote the score produced by the DNN for the booked and non-booked listings respectively
    \item $CE$ is defined as the cross entropy loss function where the input is the score difference of the booked and non-booked listing and the target label is 1
\end{itemize}

The principal metric used for offline evaluation is NDCG with binary relevance scores --- the booked listing is assigned a score of 1 and all non-booked listings have a score of 0.

\subsection{Problem Motivation}

While deep learning combined with the pairwise formulation in Equation ~\ref{pairwise_eqn} proved to be a powerful tool for optimizing offline NDCG and driving online booking gains, we soon realized that it had clear limitations. One of the most noticeable problems was the lack of diversity in search results. This issue was first brought to our attention when it was observed that for many popular destinations, the top ranked listings always seemed similar in terms of visible attributes such as price, location, capacity, and listing room type. 

\begin{figure}
\includegraphics[height=1.9in, width=3.2in]{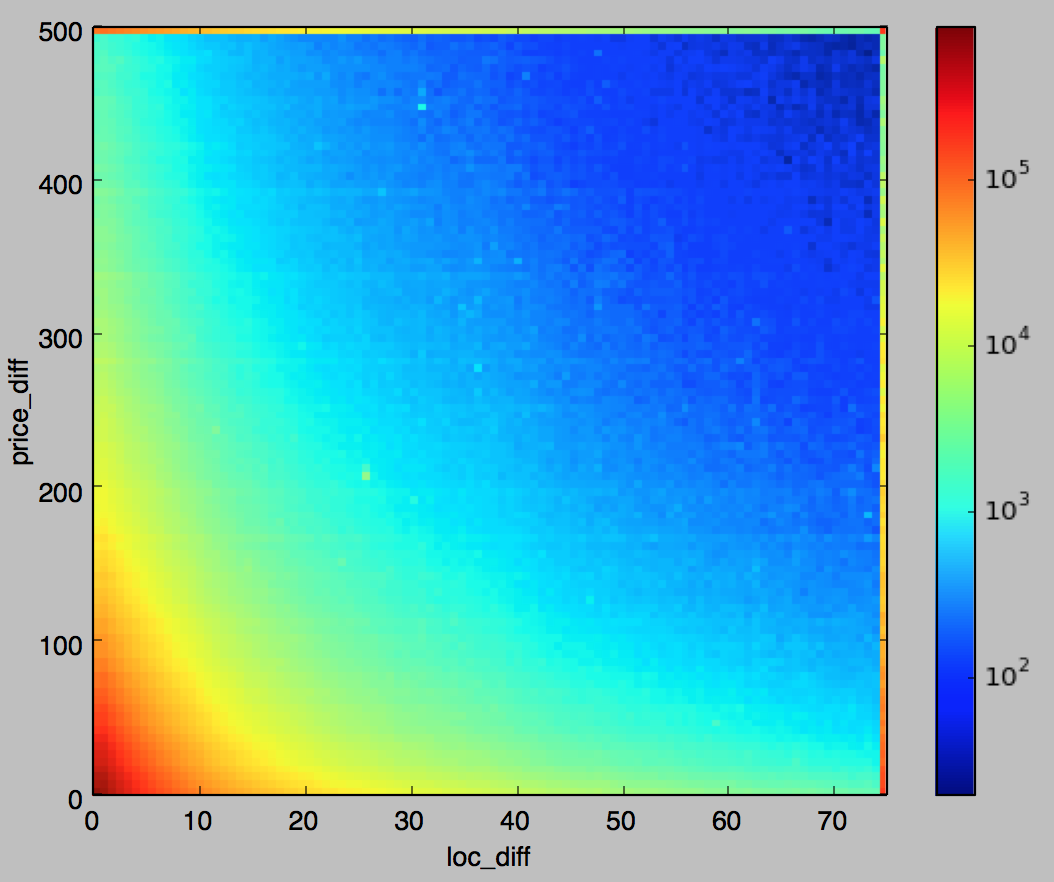}
\caption{Distribution of price and location differences from top-ranked listing pairs in search results}
\label{fig:loc_and_price_diff}
\end{figure}

This concern was further validated by looking at the data. When we sampled pairs of top ranked listings in searches (the results of which are shown in Figure ~\ref{fig:loc_and_price_diff}) we noticed a high concentration of listings with similar prices and locations. As price and location are amongst the most important factors guests use when making a booking decision, we reasoned that this lack of diversity could be detrimental for guests who did not immediately find the top search results relevant.

As it turns out, this lack of diversity was not entirely unexpected. Given this problem formulation, the DNN model was simply trained to learn what combination of features made a listing relevant for a given query and user, which implied that similar listings should have similar relevance scores. As such, the model did not have any mechanism to understand how the top ranked listings interacted with other listings shown to the user.

\subsection{Search Architecture}

The overall architecture of the search system used at Airbnb consists of two general stages. In the first stage, candidate listings matching the query parameters are retrieved and scored by the base DNN ranking model. The top $T$ of these listings (as ordered by the DNN model score) are then evaluated by what we refer to as \textit{Second Stage Rankers}. These rankers can have several purposes such as enforcing business logic or optimizing secondary ranking objectives.

\begin{figure}
\includegraphics[height=1.4in, width=3.4in]{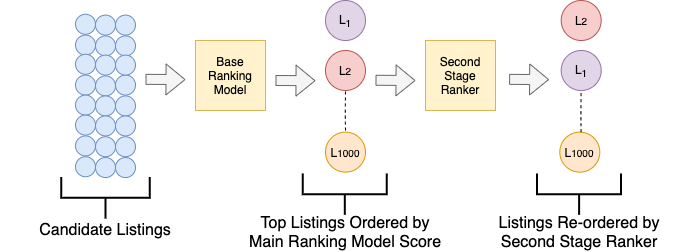}
\caption{General architecture of search system with second stage ranker}
\label{fig:search_arch}
\end{figure}

Our solutions in handling diversity generally involve building new types of second stage rankers to re-rank the top $T$ results for each search. An overall architecture diagram of this process is displayed in Figure \ref{fig:search_arch}.

\subsection{Related Work}

There exists a rich history of methods for managing diversity in ranking. Many approaches first define setwise diversity metrics as in the case of Maximal Marginal Relevance (MMR) ~\cite{carbonell1998use}. However, pure setwise metrics do not fit well in e-commerce search applications as they do not account for the significant positional bias in how results are shown to the user.

More sophisticated metrics, such as $\alpha$-NDCG ~\cite{clarke2008novelty} aim to solve this problem by modifying the definition of NDCG to include a penalty based on \textit{subtopic relevance} --- essentially rewarding items with novel subtopics while penalizing those with redundant subtopics. This framework is also challenging to directly apply to Airbnb as listings are highly variable and do not easily map in a structured way to a discrete class of subtopics. Furthermore, even if listings were mapped to subtopics, it is not entirely obvious which subtopics are relevant for a given query unlike in traditional Information Retrieval settings. 

Another class of approaches focuses on understanding diversity via adding contextual information to the model itself. These include ideas such as the listwise context model proposed in \cite{ai2018learning} and the groupwise scoring functions proposed in ~\cite{ai2019learning}. While such approaches served as important motivation, the proposed architectures did not exactly fit with our problem formulation and, particularly in the case of groupwise scoring, came with the potential for a significant increase in latency.

In summary, the main contributions made in this paper are as follows:
\begin{enumerate}
    \item Defining a diversity metric which incorporates both positional bias and a continuous distance measure between items
    \item Utilizing the distance to a target distribution as a means of measuring the diversity of the top $K$ items in a result set
    \item Providing a general method for creating a combined loss function that incorporates both relevance (via pairwise loss) and diversity (via distance to a target distribution)
    \item Creating a network architecture which combines listwise context (encoded using a Long Short-Term Memory cell) with static features to produce an embedding of the entire query context.
    
\end{enumerate}

\section{Measuring Diversity}

\subsection{Mean Listing Relevance}

In order to improve diversity, our first step was finding a concrete way to measure it. We took inspiration from the classic diversity metric MMR ~\cite{carbonell1998use} which assigned a score to each item in the result set based on a linear combination of relevance to the query and a penalty for the most similar item already selected. 

Our proposed metric differs from the traditional MMR metric in two important ways. First, we wanted to capture the idea of \textit{positional bias} since it was clear from the data that top ranked listings have a much higher chance of being booked solely due to their position. This led us to define the relevance term as the actual position discount function multiplied by the probability of a listing being booked given the query. The position discount function was empirically determined via user click-through-rates and decays in a logarithmic fashion as shown in Figure ~\ref{fig:discount_curve}. 

The second difference is that instead of using the \textit{max} as the aggregate function in the diversity term (which penalizes a new candidate based on the distance to the closest item in the current set) we use the \textit{mean}. The use of a smoother aggregation function was intended to mitigate the assumption inherent to the original MMR metric that a user is only interested in one item per category which did not generally apply in the case of Airbnb search. Our final diversity metric, which we define as \textit{Mean Listing Relevance} (MLR) is then given by: 

\begin{figure}
\includegraphics[height=1.9in, width=3.2in]{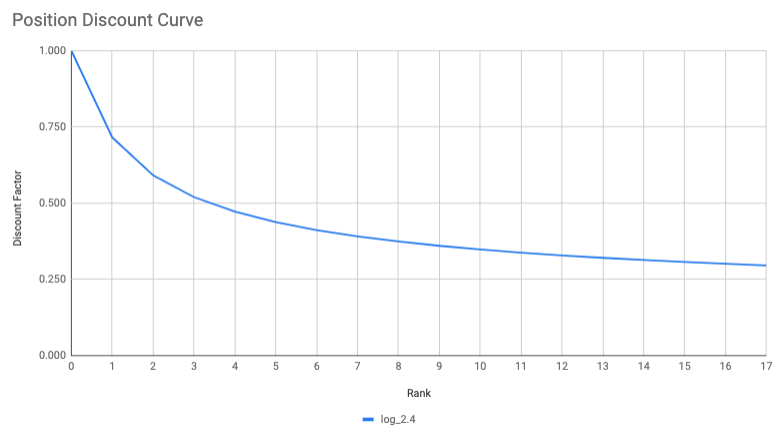}
\caption{Discount curve based on empirical click-through-rate (CTR) for each position}
\label{fig:discount_curve}
\end{figure}

\begin{equation}
\begin{aligned}
MLR(S) = \sum_{i=1}^{N} \left[ (1 - \lambda)c(i)P_{Q}(l_{i}) + \lambda \sum_{j < i} \frac { d(l_{i},l_{j})}{i}  \right] 
\end{aligned}
\label{ld_metric}
\end{equation}

Where:

\begin{itemize}
    \item $S = \{ l_{1}...l_{N} \}$ is an ordered sequence of $N$ listings
    \item $P_{Q}(l_{i})$ is the probability of listing $l_{i}$ being booked given the query $Q$
    \item $c(i)$ is the positional discount function described above
    \item $d(l_{i},l_{j})$ is an arbitrary distance measure between two listings
    \item $\lambda$ is a hyperparameter which controls the trade off between the objectives relevance and diversity
\end{itemize}

In practice, we tended to focus on listing diversity within the context of a single page which meant we usually selected $N$ to be the average number of listings displayed at a time. We also chose to use $\lambda = 0.15$ via offline analysis of how different values of $\lambda$ were correlated with other simple diversity metrics such as the variance of price, person capacity and room type of listings.

\subsection{Listing Distance Metric}

Armed with an overall definition for diversity, the next problem we faced was how to define a distance metric between listings. While this can be accomplished via Deep Learning using \textit{word2vec} type approaches ~\cite{mikolov2013distributed}, we decided to start with a representation that was easy to interpret. Specifically, we represented each listing as a vector with each element derived from tangible listing attributes such as price, location (represented as latitude and longitude), person capacity, number of bathrooms, and room type. For the continuous attributes, normalization strategies were applied as in ~\cite{kdd19} though normalization constants were derived from the specific query retrieval set rather than the global distribution. For categorical attributes, such as room type, one-hot encoding was used. 

\begin{figure}
\includegraphics[height=2.0in, width=3.4in]{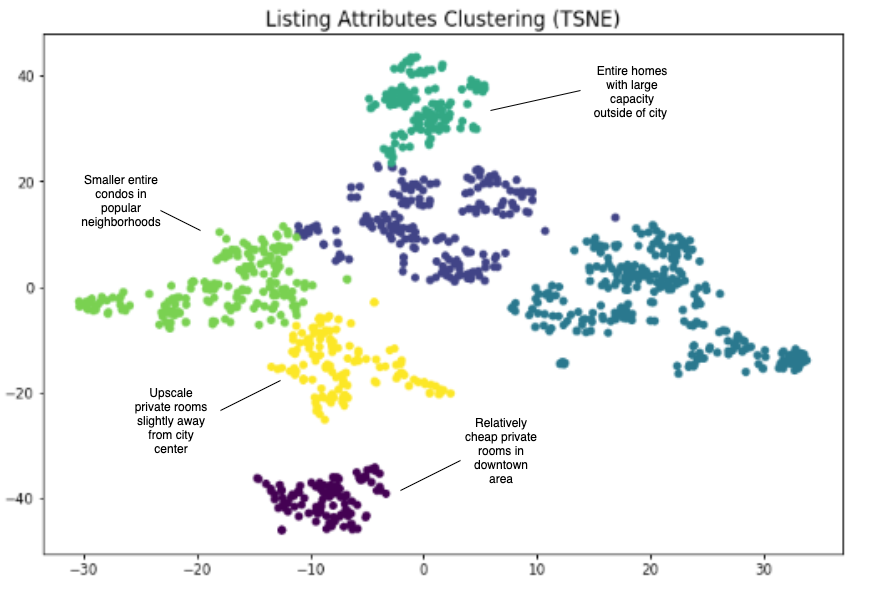}
\caption{Clustering of listings retrieved for a popular query using TSNE and K-Means (K=6). Clusters are labelled with the general type of listings they represent.}
\label{fig:tsne_listing}
\end{figure}

The plot in Figure \ref{fig:tsne_listing} shows a visualization of these listing vectors for a popular location using TSNE ~\cite{misreadtsne} to project the data into a lower dimensional space. As expected, we see the formation of clusters with human-understandable semantics such as cheaper private rooms which are close to the city center and more expensive entire homes that are further away. These clusters gave us confidence that our distance measure was encoding diversity in a reasonable way and optimizing MLR would produce a result set that was more distributed among such clusters.

\subsection{Distribution Distance as a Diversity Measure}

Our second method of measuring diversity was motivated by the particular case of location diversity which is generally one of the most visible properties of the result set when users perform a search on Airbnb. Historically, first page results for most cities had been largely clustered in the downtown core or in a single popular tourist area, which fails to capture the broad range of intents across users. For example, when searching for Orlando, FL users on family vacations might be interested in staying in the Disney World area while business travellers would prefer the city center. Ideally, the proportion of listings from each area should match the overall preference of users.

In order to capture this intuition, we chose to define diversity with respect to some attribute as the \textit{Hellinger distance} between the empirical distribution of the top $K$ results and an \textit{ideal} distribution of our choosing.

\subsubsection{Location Diversity}
\hfill\\
In the case of location diversity, the ideal distribution was constructed based on user engagement data as this was our strongest signal on what areas were relevant for a given query.

In order to build this ideal location distribution, we needed a way to aggregate user engagement signals into a geographic region. To accomplish this, we leveraged a KD-Tree \cite{bentley1975multidimensional} data structure as merging KD-Tree nodes is a relatively cheap operation. First, we divided the earth into leaf nodes of roughly equal listing density and, for a given query, recursively merged leaf nodes which fell below a certain threshold of engagement data. Once each leaf node had sufficient data, the ideal distribution was defined as the set of leaf nodes where the probability mass was weighted by the user engagement in that node relative to the total engagement data for the query similar to \cite{airbnb_locrel}. In general, this procedure had the effect of preferring roughly the same amount of listings in smaller, more popular areas (such as city centers) as in larger, less relevant areas such as those in the suburbs.

\begin{figure}
\includegraphics[height=1.9in, width=3.2in]{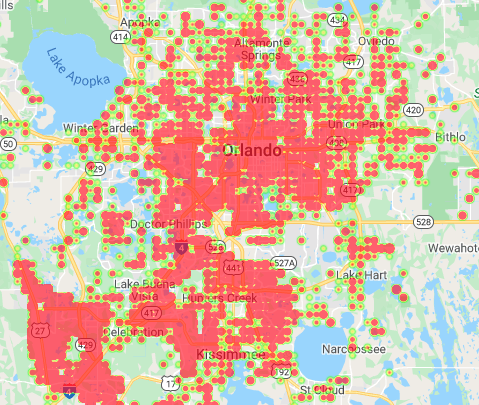}
\caption{Heatmap of user engagement data for Orlando, FL. Notice strong concentration near both the city itself and popular tourist areas such as Disney World resorts}
\label{fig:orlando_heatmap}
\end{figure}

In Figure \ref{fig:orlando_heatmap} we show a heatmap of this user engagement distribution for Orlando - notice the high concentration both within the city and popular areas further away from the city such as nearby Disney World.

\subsubsection{Price Diversity}
\begin{figure}
\includegraphics[height=1.9in, width=3.2in]{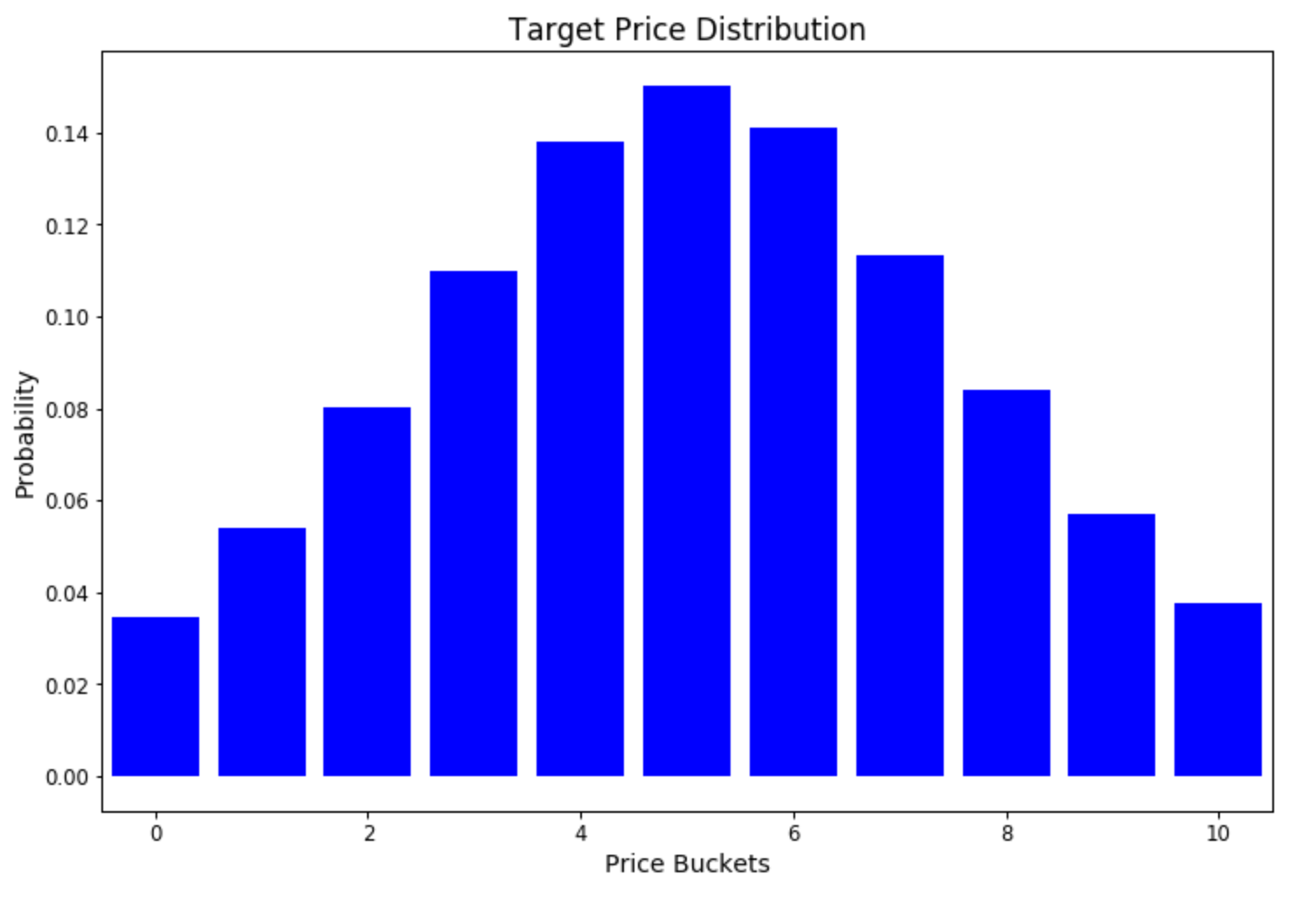}
\caption{Target price distribution. The range on the X-axis is determined by the expected price for a specific query.}
\label{fig:ideal_price}
\end{figure}

\hfill\\
Another important attribute for diversification was listing price. For the ideal price distribution, we chose a normal distribution over a range of price buckets as shown in Figure \ref{fig:ideal_price}. The motivation behind this choice was that ideally users should be able to compare listings from a wide variety of price points though the majority of them should still be close to the \textit{expected price} for a query. This expected price was computed based on a simple NN model using query features such as number of guests, number of nights and market.

In order to determine the price bucket for a given listing, min-max normalization based on an interval $[ p_{min}, p_{max} ]$ was used. The endpoints of the interval were determined by multiplying the expected price for a query by chosen constants. Specifically, we have $p_{min} = \alpha * E_{price}$ and $p_{max} = \beta * E_{price}$ where $\alpha$ and $\beta$ are hyperparameters to control the size of the price range.

\section{Methodology}

Over the past two years, several approaches have been implemented with the goal of improving search result diversity. The first family of solutions center around more heuristic systems while later attempts focus on incorporating diversity into the problem formulation itself.

\subsection{Second Stage Greedy Ranker}

Our first observation was that maximizing MLR is NP-Hard due to the sub-modular nature of Equation \ref{ld_metric} which is a fairly common property of setwise diversity functions ~\cite{nemhauser1978analysis}. As such, we decided to employ a greedy algorithm which constructed the result set incrementally by taking the best candidate listing at each stage.

\subsection{Second Stage Location Diversity Ranker}

As optimizing for all types of diversity simultaneously proved to be a challenging problem, our next approach involved focusing specifically on the important case of location diversity. To accomplish this, we crafted a loss function which would take into account both location diversity and relevance:

\begin{equation}
\begin{aligned}
Loss_{locDiv}(S) =   (1 - NDCG_{F}(S)) +  \lambda_{loc} H(L_{Q} || L_{S})
\end{aligned}
\label{loc_rel_loss}
\end{equation}

Where: 

\begin{itemize}
    \item $S = \{ l_{1}...l_{N} \}$ is a list of $N$ listings
    \item $H$ is the \textit{Hellinger distance} function between two discrete probability distributions
    \item $L_{Q}$ is the target location distribution over the  KD-Tree nodes for query $Q$
    \item $L_{S}$ is the empirical location distribution over KD-Tree nodes for listings in $S$
    \item $NDCG_{F}$ represents the NDCG of the input list under the base DNN model's ranking function $F$
    \item $\lambda_{loc}$ is a hyper-parameter that controls the importance of relevance vs. location diversity
\end{itemize}

In Equation \ref{loc_rel_loss} we are using the NDCG under $F$ as a way of measuring the deviation of the input list $S$ from the case where the listings are completely sorted in order of relevance as given by the DNN's model score. A useful property is that both the NDCG function and the Hellinger distance fall within the range $[0,1]$ which makes the value of the trade-off parameter $\lambda_{loc}$ easy to interpret. 

Unfortunately, there is no simple way to minimize the loss in Equation \ref{loc_rel_loss} using standard techniques such as gradient descent as the KD-Tree node each listing maps into is always fixed. Despite this, the change in both the NDCG and Hellinger distance functions from swapping any two listings can be computed in $O(1)$ which is an ideal property to have for non-gradient based optimization techniques such as \textit{simulated annealing} \cite{simulated_annealing}. 

Thus our solution involved building a second stage ranker which ran multiple iterations of simulated annealing  in order to construct a new set of listings. At each iteration, a random listing from the candidate set was swapped with a listing in $S$. If this swap caused the loss to decrease then it was always accepted. If the loss increased, then the swap was accepted with some probability which was a function of the loss difference and the current state of the simulated annealing algorithm.

\subsection{Second Stage Model with Combined Loss Function}

One clear limitation of the previous approaches was that they lacked the ability to make more nuanced trade-offs between relevance and diversity. Combining both objectives into a single loss function that could be optimized via gradient descent, would enable the model to learn complicated interactions between listing features, query features and diversity objectives.

The first step in our solution was devising a method to optimize a loss function for matching a target distribution. Concretely, we assume

\begin{itemize}
    \item Each training example consists of $N$ listings
    \item The target distribution is discrete and has $K$ possible values (referred to as \textit{buckets})
    \item The mapping from each listing to a bucket is always fixed
    \item The loss to be minimized is the \textit{Hellinger distance} between the target distribution and the empirical distribution formed from the top $T$ listings (where $T \leq N$). 
\end{itemize}

Our first inclination was to simply use the Hellinger Distance between the empirical and target distributions as a loss. However, we quickly found this did not work, as the bucket for each listing was constant and did not have any direct dependency on the DNN weights --- this meant there was no mechanism for backpropogation to adjust the network weights and minimize this loss directly.

In order to solve this problem, we defined a surrogate cross-entropy loss function. The main insight was that, for each bucket, we can define a binary label that indicates whether the value of the current distribution for that bucket is above or below the target value. For each listing, we use a weighted version of the cross entropy loss where the weight is proportional to the difference between the current and target value for the bucket that listing maps to.

The mechanism behind this surrogate loss is that if a listing maps into a bucket and the number of total listings in that bucket currently \textit{exceeds} the target value, then the network weights will be adjusted so as to \textit{decrease} the logit produced for the listing. This decreased score will cause the listings rank to drop below $T$ hence the number of listings in the bucket will decrease and move towards the target value.

We show an abstracted version of the TensorFlow$^{TM}$ code for this surrogate loss function construction in Table \ref{distributionloss}.

\begin{table}
\begin{lstlisting}[basicstyle=\tiny]
import tensorflow as tf
def compute_distribution_loss(top_ranked_logits, bucket_vals, target_distribution):
  '''
  Computes the distribution loss by forming a surrogate CE loss for each listing.
  top_ranked_logits contains the current logits for each listing produced by the network. Shape: [BATCH_SIZE, NUM_LISTINGS]
  bucket_vals one hot encodes the bucket each listing maps into. Shape: [BATCH_SIZE, NUM_LISTINGS, NUM_BUCKETS]
  target distribution is the ideal distribution to match. Shape: [BATCH_SIZE, NUM_BUCKETS]
  '''

  # compute empirical distribution by summing over listings
  distribution = tf.reduce_sum(bucket_vals, axis=1) / NUM_LISTINGS
  direction_mask =  tf.cast(distribution < target_distribution, tf.float64)
  direction_mask_tiled = tf.reshape(
      tf.tile(direction_mask,[1, NUM_LISTINGS]),
      shape=[BATCH_SIZE, NUM_LISTINGS, NUM_BUCKETS])
  surrogate_label = tf.reduce_sum(tf.multiply(filtered_buckets, direction_mask_tiled), axis=2)
  sqdiff = tf.squared_difference(tf.sqrt(distribution),  tf.sqrt(target_distribution))
  # weights on surrogate loss are proportional to Hellinger Distance for each bucket
  target_weights = tf.reduce_sum(
    tf.multiply(
      tf.reshape(tf.tile(tf.nn.l2_normalize(sqdiff), [1, NUM_LISTINGS]),
        shape=[BATCH_SIZE, NUM_LISTINGS, NUM_BUCKETS]),
      bucket_vals),
    axis=2)
  surrogate_loss = tf.reduce_mean(
    tf.multiply(
      tf.nn.sigmoid_cross_entropy_with_logits(labels=surrogate_label, logits=top_ranked_logits),
    target_weights))

  return surrogate_loss

\end{lstlisting}
\caption{Abstracted TensorFlow\textsuperscript{TM} code for surrogate loss for distribution matching}
\label{distributionloss}
\end{table}

With this method, we could now define an overall loss which was a linear combination of the standard pairwise loss, location distribution loss and price distribution loss. We then trained a second stage model using this combined loss function.

\subsection{Leveraging the Context}

Our final class of solutions was motivated by taking a step back to look at the problem from another angle. Ultimately, our goal was for the model to learn what results were relevant for a given query but one key piece of data that was missing was the context of the retrieval set itself. For example, it is quite possible a listing was booked because it was one of the few still available in a popular area. While diversity did explain this type of phenomenon, a more general solution would involve encoding the context of the retrieval set itself for the model to learn why this specific listing ended up being more relevant than others.

\subsection{Second Stage Model with Contextual Features}

The first attempt to incorporate contextual information led to us to create a second stage model with additional features derived from the top $K$ listings. These features included mostly hand-crafted aggregations of listing attributes such as the mean and variance of the price, location, room type, and person capacity. Our expectation was that these contextual features would enable the model to approximate when a listing was diverse relative to other listings available and if it should be up-ranked accordingly.

\subsection{Second Stage Model with Query Context Embedding}

Our final strategy involved fully embracing the adventure of deep learning as we reasoned that using hand-crafted aggregations was likely not the best way to represent the query context. Inspired from the ideas in \cite{ai2018learning}, we decided to use Recurrent Neural Networks (RNNs) to embed the listwise context of the top results and subsequently re-rank listings given this information.

We chose to use Long Short-Term Memory (LSTM) cells as they are able to better capture long term dependencies and mitigate the vanishing gradient problem when compared to vanilla RNNs \cite{hochreiter1997long}. This better performance is achieved by having the LSTM  maintain an internal cell state computed by operations on various gates. Specifically, given a sequence of listings,  the output of the LSTM at each time step is just a function of the previous hidden state and the current input:

\begin{equation}
\begin{aligned}
h_{t}, c_{t} =   LSTM_{{W_{i},W_{f},W_{o},W_{c}}}(l_{t},h_{t-1})
\end{aligned}
\end{equation}

Where:

\begin{itemize}
    \item $L = \{ l_{1} \cdots l_{N} \}$ is a sequence of $N$ listing feature vectors
    \item $h_{t-1}$ is the hidden state from the previous time step
    \item $h_{t}$ and $c_{t}$ are the hidden state and cell output at time $t$
    \item $W_{i},W_{f},W_{c},W_{o}$ are the learnable parameters for the input gate, forget gate, output gate, and cell state respectively
\end{itemize}

Thus, we can think of $h_{N}$, the final hidden state of the LSTM after feeding in $N$ listings, as a \textit{sequence embedding} which summarizes the listwise context.

\subsubsection{System Architecture}
\hfill\\
Once we had encoded the listwise context using a LSTM, we could then view the problem as one of re-ranking the listings given this sequence embedding. This led to a modified ranking system architecture where $L$, $Q$, and $U$ denote the feature vectors for the listing, query and user respectively.

\begin{enumerate}
    \item Score each candidate listing using our default DNN ranking model, $F$, by computing a score $F(L,Q,U)$
    \item Sort candidate listings by the default model score, retaining the top $T\approx1000$ listings as in Figure \ref{fig:search_arch}
    \item Compute the sequence embedding $h_{N}$, by feeding the top $N$ listings into the LSTM
    \item Use a second-stage DNN model, $H$, to compute a new score given the sequence embedding, $H(L,Q,U, h_{N})$, and re-rank the top $K$ results by this score.
\end{enumerate}

One of the main design choices was the value of the hyperparameters, $N$, the size of the input sequence to the LSTM and, $K$, the number of results to re-rank. Theoretically, we would like $N$ to be as large as possible since longer sequences should contain more information. However, in reality, there are practical considerations as using a larger $N$ directly increases the size of both the raw logs and training data. 

At training time, only listings shown to the user contributed to the pairwise loss though up to $N$ listings were used to create the sequence embedding. This was done in order to remain faithful to the pairwise problem formulation by not treating unshown listings as negative examples. Subsequently, this made the choice of $K$ more complicated since if $K$ was too small we may not be re-ranking enough results to produce a measurable effect. On the other hand, if $K$ was too large then the online implementation would deviate significantly from the training objective. Via offline simulation, we empirically determined a value of $K$ which achieved a reasonable trade-off between these constraints.

\subsubsection{Network Architecture}
\hfill\\
The architecture for the new second stage DNN model, $H$, was inspired by the two-tower architecture discussed in \cite{improvingdeeplearning}. The main takeaway is that separating the listing independent and dependent features into two separate sub-networks increased the models ability to generalize.

\begin{figure}
\includegraphics[height=1.9in, width=3.2in]{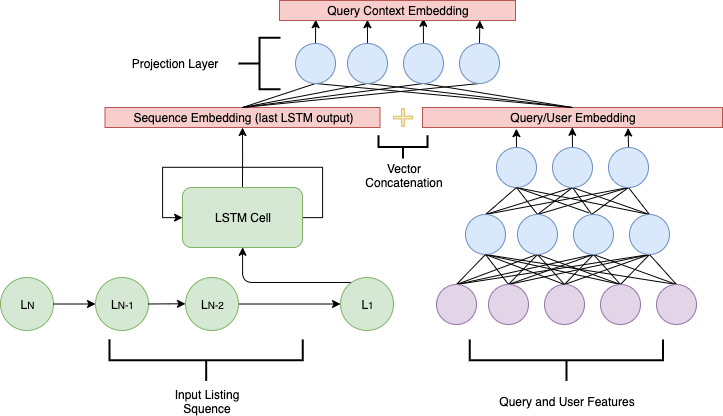}
\caption{Architecture to compute embedding of given query and input listing sequence}
\label{fig:egghead_arch}
\end{figure}

Extending this idea, we concatenated the LSTM output $h_{N}$ with the output of the listing independent sub-network (fed by query and user features) and added a final projection layer to produce a vector that represented the ideal listing given a specific query and listwise context. This architecture is shown in Figure \ref{fig:egghead_arch} and we refer to the output of this network as a \textit{query context embedding}. Abstracted TensorFlow code to compute this query context embedding is shown in Table \ref{querycontextembedding}. 

\begin{table}
\begin{lstlisting}[basicstyle=\tiny]
import tensorflow as tf
def apply_fc_layer(feature_vec, weights, biases):
  '''
  Compute one fully connected layer by applying weights and biases
  '''
  return tf.nn.tanh(tf.matmul(features, weights) + biases)

def compute_tower(feature_vec, w0, b0, w1, b1):
  ''' 
  Compute sub-tower - two layer DNN
  '''
  h1 = apply_fc_layer(tensor, w0, b0)
  h2 = apply_fc_layer(tensor, w1, b1)
  return h2

def compute_query_context_embedding(listing_features_seq, query_features, seq_lengths):
  '''
  Compute query context embedding given sequence of listing features, query features
  listing features has shape [batch_size, # of listings per example, # of listing features]
  seq_lengths has shape [batch_size, # of visible listings in batch]
  '''
  query_tower_output = compute_tower(query_features, query_w0, query_b0, query_w1, query_w2)
  output, state = tf.nn.dynamic_rnn(lstm_cell, listing_features_seq, sequence_length=seq_lengths)
  sequence_embedding = state.h
  query_and_rnn_concat = tf.concat([query_tower_output, sequence_embedding], -1)
  query_context_embedding = apply_fc_layer(query_and_rnn_concat, proj_weights, proj_biases)
  return query_context_embedding

\end{lstlisting}
\caption{Abstracted TensorFlow\textsuperscript{TM} code for computing query context embedding}
\label{querycontextembedding}
\end{table}

As in \cite{improvingdeeplearning}, the euclidean distance between the listing embedding and the query context embedding was used to measure how close a candidate listing was to the ideal listing given the query and list-wise context. Similar to the original problem formulation, we used a pairwise loss function which has the effect of moving the booked listing embedding closer to the query context embedding while pushing non-booked listings away.

\begin{figure}
\includegraphics[height=1.9in, width=3.2in]{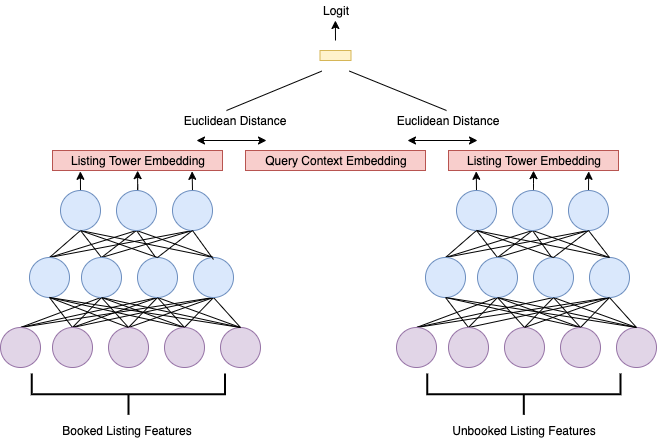}
\caption{Overall architecture of second-stage re-ranking model. The query context embedding is generated from the network in figure \ref{fig:egghead_arch}}
\label{fig:overall_arch}
\end{figure}

\section{Experiments and Discussion}

In Table \ref{ld_and_ndcg_metric_table} we report the percent difference in MLR and NDCG for the various approaches described previously. In addition, the percent difference in price and location diversity metrics --- measured by the distance to ideal distributions are shown in Table \ref{ld_and_ndcg_metric_table}. All metrics were computed with respect to the baseline of only using the default DNN ranking model and obtained via offline simulation on test data.

We now discuss the results obtained via online A/B tests for each of the methods in more detail.

\begin{table}[h!]
\centering
\begin{tabular}{|c | c | c | c||} 
 \hline
 Method & MLR &  NDCG  \\ [0.5ex] 
 \hline
 \hline
 Greedy Ranker & 2.45\% & $-$0.96\%  \\
 \hline 
 Location Diversity Ranker & 1.04\% & $-$0.13\%  \\
  \hline
 Contextual Features & 0.56\% & 0.21\%  \\
  \hline
 Combined Loss Function & 1.89\% & 0.03\%  \\
 \hline
 \textbf{Query Context Embedding} & \textbf{1.97\%} & \textbf{1.26\%}  \\
 \hline
\end{tabular}
\caption{Percent difference in MLR and NDCB for various second stage diversity rankers. Baseline is default DNN model.}
\label{ld_and_ndcg_metric_table}
\end{table}

\begin{table}[h!]
\centering
\begin{tabular}{|c | c | c | c||} 
 \hline
 Method & Location Diversity & Price Diversity \\ [0.5ex] 
 \hline
 \hline
 Greedy Ranker & 2.08\% & 1.13\%  \\
 \hline 
 Location Diversity Ranker & 4.61\% & 0.87\%  \\
  \hline
 Contextual Features & 0.65\% & $-$0.35\%  \\
  \hline
 \textbf{Combined Loss Function} & \textbf{2.63}\% & \textbf{1.81}\%  \\
 \hline
 Query Context Embedding & 2.42\% & 1.16\%  \\
 \hline
\end{tabular}
\caption{Percent difference in location diversity and price diversity for various second stage diversity rankers using the distance from ideal distribution method. Baseline is default DNN model.}
\label{ld_metric_table}
\end{table}

\subsection{Second Stage Greedy Ranker}
The first set of online experiments related to diversity involved applying the second stage greedy ranker for optimizing MLR. Unfortunately, we observed slightly negative or neutral results in all iterations. One clear issue was the inability of this ranker to capture more complicated patterns with respect to diversification. For example, seeing diverse listings might be useful at the start of a users journey to help them understand the breadth of available choices but is less important as their preferences begin to narrow.

The end result was that in most cases we seemed to be over-diversifying results and creating friction for our users. This was demonstrated by a statistically significant increase in metrics such as price filter usage and listing views --- all indications that users had to work harder to find what they wanted.

\subsection{Second Stage Location Diversity Ranker}

In terms of offline metrics, the location diversity ranker was quite promising. We were able to significantly decrease the distance to the ideal location distribution while only incurring a slight hit in NDCG. We observed positive results during the online test as well --- the highlight being a statistically significant gain in bookings from new users (+1\%).

In addition, we also saw a statistically significant increase for bookings in countries such as China (+3.6\%) largely owing to the fact that, prior to this re-ranker, a simple heuristic which promoted listings near the city center was used. Clearly, by taking a broader view of location relevance, users were exposed to listings in popular areas nearby the city which resulted in a positive experience.

\begin{figure}
\includegraphics[height=1.0in, width=3.4in]{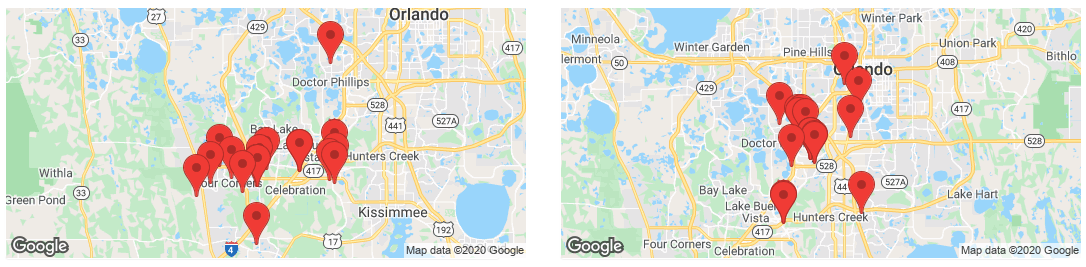}
\caption{Example of location distribution before (left) and after (right) location diversity re-ranker applied.}
\label{fig:orlando_sxs}
\end{figure}

As a final sanity check, we inspected the  motivating example of Orlando, FL and were pleased to find cases where after applying this location diversity re-ranker the result set was far more distributed over relevant areas as opposed to being clustered together. An example of this is shown in Figure \ref{fig:orlando_sxs}.

This location diversity ranker was launched to production where it remained for over 2 years before being replaced by the more complex model based approaches.

\subsection{Combined Loss Function}
When combining diversity and relevance objectives into a single loss function for the model, the main benefit compared to the non-model based approaches was the ability to directly improve diversity without degrading NDCG.

However, we were faced with fairly neutral results during the online A/B tests. The most likely explanation seemed to be a statistically significant increase in the proportion of listings within extreme price ranges --- both very cheap and very expensive. In particular, we hypothesized that increasing the proportion of listings that were expensive led to a poor user experience especially for new guests who are price conscious. This hypothesis was supported by the fact that we observed a statistically significant decrease in new guest bookings.

While we iterated on the shape of the ideal price distribution and how each term in the overall loss function was weighted, we found it difficult to fine-tune the behavior of the model. This highlighted another trade off between model based approaches as opposed to the simple heuristic strategies used in previous attempts --- it was much harder to reason about the final effect when adjusting design parameters.

\subsection{Second Stage Model with Contextual Features}
The second stage model with contextual features added showed a modest gain in NDCG but it was less than the typical gain required to see a measurable effect in online experiments. Furthermore, we observed only slight changes in all diversity metrics. Our general conclusion was that these hand crafted contextual features alone were not powerful enough to move the needle in terms of both diversity and relevance.

\subsection{Second Stage Model with Query Context Embedding}
The approach with the most promising offline results was the second stage model with the query context embedding. We observed a significant offline gain in NDCG (the highest relative to all other approaches) along with solid improvements in all diversity metrics. This suggested to us that by encoding the listwise context the model was able to better recognize situations where diversity should be applied.

During online testing, we observed several positive results that tracked our offline metrics quite closely. These included a statistically significant increase in online NDCG (+1.2\% --- one of the largest in the past few years), overall bookings (+0.44\%) and in bookings from new guests (+0.61\%). Due to these strong results, this model was launched to production.

Interestingly, the improvement in our diversity metrics (especially price diversity) were not as pronounced as in the combined loss function approach which supported our hypothesis that we may have been over-diversifying results before.

\begin{figure}
\includegraphics[height=1.9in, width=3.2in]{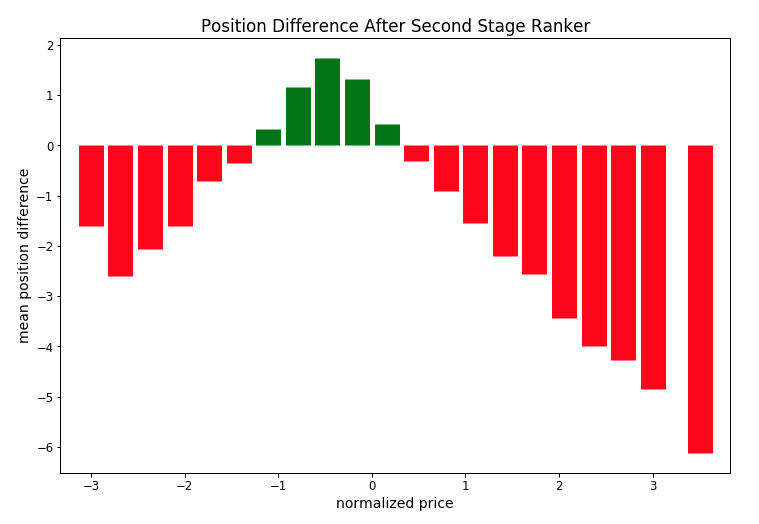}
\caption{Plot of mean rank change versus normalized price of a listing (relative to top $N$ results) when using second stage model with query context embedding}
\label{fig:pos_analysis_egghead}
\end{figure}

One lingering question we had, as DNN models are generally not very interpretable, was how to gain deeper insight into the model's behavior. To investigate this, we analyzed how the position of a listing changed as a function of its normalized price relative to the top $N$ listings after this second stage ranker was applied (shown in Figure \ref{fig:pos_analysis_egghead}).

In general, we saw that the model tended to demote listings with more extreme prices in a non-symmetric way (penalizing expensive listings more harshly). This makes sense from a diversity perspective in that once there is already a rather expensive listing in the top results the incremental value of adding another expensive listing is fairly low. As our default ranking model had the effect of consistently showing similarly priced listings together (as evidenced by the plot in Figure \ref{fig:loc_and_price_diff}), this suggests our second stage model was able to learn a correction on when this behavior was detrimental at least in the context of price.

\section{Future Work}

Though we achieved our initial goal of increasing diversity and showing more relevant results to our users, there still remain many unexplored frontiers. One promising idea is to investigate techniques such as \textit{transfer learning} to reuse layers or sub-networks from the base ranking model in the second stage model. This would allow us to both train models more efficiently and create a more robust system overall.

In terms of model architecture, there are several exciting developments  when it comes to processing sequential data. These include techniques such as attention mechanisms \cite{bahdanau2014neural} and transformer networks \cite{attention}. Lastly, we would like to explore full sequence-to-sequence models, where the output list is constructed incrementally using \textit{pointer network} architectures such as in \cite{bello2018seq2slate}. Our hope is that using more sophisticated techniques will enable the model to learn more nuanced properties of the input sequence, thereby re-ranking results in a more optimal way.

\section{Conclusion}

Our journey to understand diversity has led us down many different paths --- starting from non-model based solutions based on classic ideas, to devising a combined loss function, and culminating with a new architecture that encodes listwise context in the model itself. 

Along the way, we have learned many lessons, most of which center around the theme of thinking deeply about the trade off between various approaches. We found that non-model based approaches work well when targeting diversity along a single attribute and are relatively easy to interpret --- however, they are unable to make more nuanced trade offs and almost always sacrifice relevance. In contrast, while model based approaches are generally more difficult to reason about they offer much more powerful generalizations when provided with the right information.

Most of all, we are grateful for the endless inspiration and guidance provided from both our colleagues and the deep learning community throughout this process and look forward to new challenges on the horizon.
